# Numerical Study on Droplet Evaporation and Propagation Stability in Normal-temperature Two-phase Rotating Detonation System

Haocheng Wen, Wenqi Fan, Sheng Xu and Bing Wang[*]

School of Aerospace Engineering, Tsinghua University, Beijing, China, 100084

**Abstract:** A numerical study is carried out on the droplet-laden two-phase rotating detonation wave (RDW) of kerosene/oxygen-enriched air at normal temperature. Two types of combustors without and with the inlet mixing section (IMS) are constructed to illustrate the effect of IMS on the combustion characteristics of two-phase RDW. The important role of the preheating zone in the IMS after the back-propagation shock on the droplet evaporation is analyzed. The parameter sensitivity of RDW propagation stability to the average droplet diameter $d_0$ is further discussed. Results show that the droplets mainly evaporate after the detonation front in the combustor without IMS, and the reaction heat release is completed in a short distance, which propels continuous propagation of the detonation wave. When $d_0$ gradually increases, the droplet evaporation distance increases, and the coupling between the incident shock and reaction is continuously weakened, finally resulting in the detonation quenching. In the combustor with IMS, a preheating zone is induced close to the contact surface by the back-propagation shock of the RDW. A large number of droplets evaporate in this zone, and generate sufficient mixture of fuel vapor and oxidizer in front of detonation wave to maintain the detonation propagation. Priority to the combustor without IMS, the droplet evaporation relies less on the inlet high-temperature airflow with the assistance of preheating zone, and thus the wave propagation stability can be enhanced and the RDW can sustain for a wider range of $d_0$. The present analysis provides a new understanding of two-phase rotating detonation systems.

**Keywords**：two-phase rotating detonation, normal temperature, droplet evaporation, stability, numerical simulations

*Corresponding author: wbing@tsinghua.edu.cn

**Nomenclature**

| | |
|---|---|
| $\varphi_t$ | Spray total equivalence ratio |
| $\varphi_{pre}$ | Pre-evaporated equivalence ratio |
| $\varphi_{lb}$ | Lower detonative limit |
| $B_M$ | Mass transfer coefficient |
| $d$ | Droplet diameter |
| $\dot{m}$ | Mass flow rate |
| $P$ | Pressure |
| $T$ | Temperature |
| $Y$ | Species mass fraction |
| $u$ | velocity |
| Pr | Prandtl number |
| Re | Reynold number |
| Nu | Nusselt number |
| Sh | Schmidt number |
| Sc | Sherwood number |

*Subscript*

| | |
|---|---|
| 0 | Initial or stagnation status |
| d | Droplet parameter |
| D | Detonation parameter |
| k | Index of species |

## 1 Introduction

Rotating detonation is an important approach to realize pressure-gained combustion, which can help to improve the performance of propulsion devices based on isobaric combustion cycles. Since Voitsekhovskii [1] discovered the rotating detonation phenomenon for the first time in the 1950s and Nicholls et al. [2] completed the first attempt on the rotating detonation engine in the 1960s, the fundamental issue and engineering applications of rotation detonation have attracted extensive studies.

Based on numerous experimental and numerical studies on gaseous rotating detonation wave (RDW) with $H_2$ [3–6], $CH_4$ [7,8] and other gaseous fuels, the premilitary understanding are obtained on the propagation regimes of RDW (such as the one wave, counter two waves, longitudinal combustion, etc.) [9–12], the effects of multiple factors (such as the combustor geometry, injection configuration, and reactant properties, etc.) on detonation propagation [6,13–16], etc. These achievements have laid a foundation for multiple engineering applications of rotating detonation engines [17–19].

However, compared to gaseous fuels such as $H_2$ and $CH_4$, the liquid hydrocarbon fuels including kerosene have higher density and heating values per unit volume. Therefore, kerosene is still the best choice in the field of aviation, and it is inevitable to adopt kerosene for the future rotating detonation engines. Nevertheless, the liquid hydrocarbon fuel has to undergo spray processes including atomization, evaporation and mixing before reaction. In the rotating detonation combustor, there may also be a complicated coupling between the spray processes and the detonation. Additionally, the kerosene has lower activity than gaseous fuels. The experimental results show that the detonation cell size of gaseous kerosene/air mixture at 353 K is approximately 60 mm [20], which is several times the cell size of $H_2$/air under similar conditions [21]. These multiple factors have negative effects on the initiation and stable combustion of rotating detonation.

Previous experimental studies have shown that it is difficult to obtain the stable and self-sustaining droplet-laden two-phase RDW using kerosene/air reactants at normal temperature [22,23]. In order to promote the propagation stability of RDW, it is important to accelerate the fuel droplet evaporation by preheat the reactant or increase the fuel equivalent pre-evaporation. For example, the partially premixed mixture of preheated Jet-A (160 °C) and air (100 °C) was applied to improve the pre-evaporation degree in the experiments of Wolanski et al. [24]. Li et al. [25] preheated the supply air to 100 °C by an electronic heating device to achieve a pre-evaporation degree of 10%, and obtained the stable Jet-A/air RDW. The gaseous additive is also a common approach to increase the fuel equivalent pre-evaporation. Kindracki [22] added a small amount of $H_2$ in the inlet reactant to achieve the self-sustained RDW and investigate the ignition and propagation process. Another experiment with addition of $H_2$ or (CO + $3H_2$) was conducted by Bykovskii et al. [26], and the multi-wave rotating detonation and counter two-wave rotating detonation were obtained in their study.

The further understanding of the influence mechanism of spray parameters on the propagation stability of RDW are gained through numerical methods. By using the Eulerian–Lagrangian method, Ren and Zheng [27] numerically studied the working limit of kerosene two-phase RDW in the range of total pressure of 3~9 atm and total temperature of 900~1200 K. Hayashi et al. [28] applied the Eulerian-Eulerian method to investigate the effect of pre-evaporation degree and droplet diameter (1~10 $\mu$m) on JP-10 RDW. The detonation was found to be quenched with the droplet diameter greater than 4 $\mu$m and the pre-evaporation degree less than 20% at normal temperature. The droplet dynamics and evaporation process were further discussed in the simulations of Meng et al. [29,30] under different pre-evaporation degree, droplet diameter and inlet gas temperature (300~500 K) for n-heptane RDW.

In the two-phase rotating detonation systems studied in the above experiments and numerical simulations, the stable rotating detonation either relies on the high-temperature atmosphere to accelerate the droplet evaporation rate in the refilled zone, or relies on the pre-evaporation to make up for insufficient evaporation. However, multiple experimental studies have shown that the oxygen-enriched air can also promote the formation of two-phase RDW without any pre-evaporation at normal temperature atmosphere [31–33]. Bykovskii et al. [31] adopted kerosene/oxygen-enriched air to achieved stable two-phase RDW in a 306-mm-diameter combustor. Wang et al. [32] adopted a mixture of ($O_2$ + 1.5$N_2$) in a 100-mm-diameter combustor and studied the effect of combustion width (32 mm, 26 mm and 20 mm) on the propagation regimes of kerosene RDW. The oxygen-enriched air was also widely applied in many other experimental studies on two-phase RDW [33,34].

These experimental evidences indicate that the mechanism of droplet evaporation and detonation stability for the normal-temperature kerosene/oxygen-enriched air two-phase rotating detonation system is distinguished from the aforementioned ones. In order to explain the related mechanism, two types of combustors with and without the inlet mixing section (IMS) are proposed in this study. The different rotating detonation flow field structures are comparatively studied. The characteristics of the preheating zone in the combustor with IMS are introduced. Furthermore, the droplet evaporation mechanisms for the two combustors are analyzed. The significant role of the preheating zone in the droplet evaporation process is found for the normal-temperature rotating detonation system. Finally, we analyze the parameter sensitivity of the detonation propagation stability to the droplet size for the two combustors, and obtain the key characteristics such as detonation velocity deficit and stability limit. The combustor with IMS illustrates the working mechanism of the normal-temperature kerosene/oxygen-enriched air two-phase RDW.

## 2 Numerical Method

The droplet-laden two-phase rotating detonation phenomenon involves multiphase flow, evaporation, combustion and complicated shock waves systems. The Eulerian-Lagrangian method is adopted for this reacting multiphase flow system. The chemical reaction is assumed to take place only in the gas phase.

For fuel droplets, two-way coupling model is employed. Fuel spray is assumed as a cluster of individual spherical droplet with a distribution of droplet sizes. The collision, coalescence and condensation effects are neglected.

### 2.1 Governing equations for continuous phase

The Navier-Stokes equations for continuous phase can be written as

$$\frac{\partial \rho}{\partial t} + \nabla \cdot (\rho \mathbf{u}) = S_M \tag{1}$$

$$\frac{\partial}{\partial t}(\rho \mathbf{u}) + \nabla \cdot (\rho \mathbf{u}\mathbf{u}) + \nabla p = \nabla \cdot \boldsymbol{\tau} + \mathbf{S}_F \tag{2}$$

$$\frac{\partial}{\partial t}(\rho E) + \nabla \cdot (\rho E \mathbf{u} + p \mathbf{u}) = \nabla \cdot (\boldsymbol{\tau} \cdot \mathbf{u}) + \nabla \cdot \mathbf{q} + S_E \tag{3}$$

$$\frac{\partial}{\partial t}(\rho Y_k) + \nabla \cdot (\rho Y_k \mathbf{u}) = \nabla \cdot (\rho D_k \nabla Y_k) + \dot{\omega}_k + S_{Y_k} \tag{4}$$

where $\rho$, $p$ and $T$ are the density, pressure, and temperature, respectively. $\mathbf{u}$ is the velocity, $Y_k$ is the mass fraction of species $k$, $\boldsymbol{\tau}$ is the viscosity shear stress tensor, $E$ is the total energy which is the sum of internal energy and kinetic energy of all components,

$$E = \sum_{k=1}^{N_s} Y_k \left( \int_{T_0}^{T} c_{v,k}(\bar{T}) \mathrm{d}\bar{T} + h_{f,k} \right) + \frac{1}{2}\mathbf{u}^2 \tag{5}$$

where $c_{v,k}$ is the heat capacity of species $k$, and $h_{f,k}$ is the enthalpy of formation at a reference temperature $T_{\text{ref}}$ of species $k$. $\dot{\omega}_k$ is the combustion source term calculated by the finite reaction rate model.

The source terms $S_M$, $\mathbf{S}_F$, $S_E$ and $S_{Yk}$ represent the sources of mass of gas phase, momentum, energy, and mass fraction of species $k$, respectively. These sources are produced by discrete phase and can be written as

$$S_M = -\frac{1}{\Delta A}\sum_{N_d} \dot{m}_d \tag{6}$$

$$\mathbf{S}_F = -\frac{1}{\Delta A}\sum_{N_d}(\mathbf{F}_d + \dot{m}_d\mathbf{u}_d) \tag{7}$$

$$S_E = -\frac{1}{\Delta A}\sum_{N_d}\left(Q_d + \dot{m}_d\left(\frac{\mathbf{u}_d^2}{2} + h_{vapor}\right)\right) \tag{8}$$

$$S_{Y_k} = \begin{cases} -\dfrac{1}{\Delta A}\sum_{N_d}\dot{m}_d & ,\text{fuel} \\ 0 & ,\text{other species} \end{cases} \tag{9}$$

where $\Delta A$ is the volume of the single cell, $N_d$ is the number of droplets in the cell, $\dot{m}_d = \mathrm{d}m_d/\mathrm{d}t$ is the mass variation rate of all droplets in the cell, $\mathbf{u}_d$ is the droplet velocity, $\mathbf{F}_d$ is the drag force of droplet, $Q_d$ is the heat convection from the gas phase to droplets, and $h_{vapor}$ is the enthalpy of the fuel vapor.

Additionally, the ideal gas equation of state is assumed,

$$p = \rho R_u T \sum_{k=1}^{N_s} \frac{Y_k}{W_k} \tag{10}$$

where $W_k$ is the molecular weight of species $k$, $R_u$ is the universal gas constant.

**2.2 Governing equations for discrete phase**

Fuel spray is assumed as a distribution of discrete fuel droplets. The individual droplet is released and traced in the Lagrangian trajectory model. In this study, the discrete droplets are assumed as points, and the governing equations are [35]

$$\frac{\mathrm{d}\mathbf{x}_d}{\mathrm{d}t} = \mathbf{u}_d \tag{11}$$

$$\frac{\mathrm{d}\mathbf{u}_d}{\mathrm{d}t} = \frac{\mathbf{F}_d}{m_d} = \left(\frac{f_1}{\tau_d}\right)(\mathbf{u}_{@d} - \mathbf{u}_d) \tag{12}$$

$$\frac{\mathrm{d}T_d}{\mathrm{d}t} = \frac{Q_d + \dot{m}_d L_V}{m_d c_L} = \left(\frac{f_2}{\tau_d}\right)\left(\frac{\mathrm{Nu}}{3\mathrm{Pr}}\right)\left(\frac{c_d}{c_L}\right)(T_{@d} - T_d) + \left(\frac{\dot{m}_d}{m_d}\right)\frac{L_V}{c_L} \tag{13}$$

$$\frac{\mathrm{d}m_p}{\mathrm{d}t} = -m_p\left(\frac{1}{\tau_d}\right)\left(\frac{\mathrm{Sh}}{3\mathrm{Sc}}\right)\ln(1 + B_M) \tag{14}$$

where $\mathbf{x}_d$ is the droplet position, $\mathbf{u}_{@d}$ and $T_{@d}$ are the gas velocity and temperature seen

by the droplets at its position, $T_d$ is the droplet temperature, $m_d$ is the droplet mass, $\tau_d$ is the characteristic relaxation time of the droplet, $c_L$ is the specific heat of the droplet, $L_V$ is the latent heat of evaporation, $f_1$ is the modified drag coefficient, $f_2$ is modified coefficient for convection heat transfer in evaporation, and $B_M$ is the mass transfer coefficient.

The characteristic relaxation time of the droplet $\tau_d$ can be written as

$$\tau_d = \frac{\rho_d d_d^2}{18\mu} \tag{15}$$

where $\mu$ is the kinetic viscosity coefficient.

Four non-dimensional parameters, Pr (Prandtl number), $Nu_d$ (Nusselt number), Sh (Sherwood number), and $Sc_d$ (Schmidt number), are formulated as [35]

$$\Pr = \frac{\mu c_p}{\lambda}, Nu_d = 2 + 0.552 \operatorname{Re}_d^{0.5} \Pr^{1/3}$$
$$Sc = \frac{\mu}{\rho D}, Sh_d = 2 + 0.552 \operatorname{Re}_d^{0.5} Sc^{1/3} \tag{16}$$

where $\operatorname{Re}_d$ is the Reynold number of droplets

$$\operatorname{Re}_d = \frac{\rho |\mathbf{u}_{@d} - \mathbf{u}_d| d_d}{\mu} \tag{17}$$

The modified drag coefficient $f_1$ is the function of $\operatorname{Re}_d$ [36]

$$f_1 = \frac{\operatorname{Re}_d}{24}\left(\frac{24}{\operatorname{Re}_d}\left(1 + 0.15\operatorname{Re}_d^{0.687}\right) + \frac{1}{1 + 42500\operatorname{Re}_d^{-1.16}}\right), \operatorname{Re}_d \leq 2\times 10^5 \tag{18}$$

The modified coefficient for convection heat transfer in evaporation $f_2$ can be written as a function of dimensionless evaporation coefficient [35]

$$f_2 = \frac{\beta}{e^\beta - 1} \tag{19}$$

where $\beta$ is the dimensionless evaporation coefficient with the definition as

$$\beta = -1.5 \Pr \tau_d \frac{\dot{m}_d}{m_d} \tag{20}$$

The validation of the droplet evaporation model used in the study can refer to Appendix A.

**2.3 Reaction Mechanism**

The reactant employed in the study are kerosene (substituted by $C_{10}H_{20}$) and oxygen-enriched air ($O_2$ + 1.5$N_2$) at the normal temperature. A two-step reaction mechanism [37] for kerosene/oxygen combustion is adopted

$$\begin{aligned} &KERO+10O_2 \rightarrow 10CO+10H_2O \\ &CO+0.5O_2 \rightleftharpoons CO_2 \end{aligned} \quad (21)$$

The reaction rate $k_{f,1}$ and $k_{f,2}$ are modified with equivalent ratio

$$\begin{aligned} k_{f,1} &= A_1 b_1(\varphi) e^{(-E_{a,1}/RT)} [KERO]^{n_{KERO}} [O_2]^{n_{O_2,1}} \\ k_{f,2} &= A_2 b_2(\varphi) e^{(-E_{a,2}/RT)} [CO]^{n_{CO}} [O_2]^{n_{O_2,2}} \end{aligned} \quad (22)$$

where $A_i$ ($i$ = 1,2) are pre-exponential coefficients, $E_{a,i}$ ($i$ = 1,2) are activation energy, and $b_i(\varphi)$ are correction coefficients which can refer to Reference [37].

**2.4 Numerical schemes**

For the solution of gas phase governing equations, the finite difference method (FDM) is used. The high order scheme is used to capture the shock wave structure in the flow field in this study. For the convection term, the WENOLF scheme developed by Hu et al. [38] is applied. For the viscous diffusion term, this study adopts the sixth-order central difference scheme. The third-order Runge-Kutta algorithm is used to solve the solution of the governing equations.

For the solution of discrete phase governing equations, the second order Admas-Bashforth scheme is used for the time discretization. The gas variables at the droplet location are obtained by the fourth order Lagrangian interpolation method. The above methods has been validated in Reference [27,39] and used for problems on the compressible reacting flow.

**3 Computation Set-up**

*3.1 Two Rotating Detonation Combustor Models*

The two-dimensional (2D) unrolled rotating detonation combustors are simulated in the study. As shown in Figure 1a, the dimension of the computational domain is 0.2 m in the *x*-direction (the axial direction of combustor) and 0.45 m in the *y*-direction (the circumferential direction of combustor). Based on the grid sensitivity analysis (see Appendix B), the uniform grid size is chosen as 0.25 mm and the mesh size is 800×1800. The time step size is taken as 0.01 $\mu$s and it has been found that further decreasing the time step size has little effect on the numerical results. The top boundary ($x = 0$ m) is the combustor inlet and refers to the variable cross-section inlet model given by Fievisohn and Yu [40]. The total pressure loss caused by the channel expansion is considered in this model. The expansion ratio is given as $A_1/A_3 = 8.8$ in the study. The bottom side ($x = 0.2$ m) is the outlet boundary with ambient pressure of 100 kPa. The right ($y = 0$ m) and left side ($y = 0.45$ m) is the periodic boundaries. In the following cases, the detonation wave generally propagates from right to left.

In the real rotating detonation combustors, the oxidizer and liquid fuel are usually mixed within a certain distance downstream of the inlet boundary to reach the flammability limit. In order to study the influence of the mixing section on the droplet evaporation and detonation propagation behavior, two types of combustors are constructed in the study, namely the Combustor A (Figure 1a) without the inlet mixing section (IMS) and Combustor B with IMS (Figure 1b). As shown in Figure 1b, the IMS is simulated by closing the chemical reaction in the zone of $x < x_{\text{mix}}$. The reaction is stopped at $x = x_{\text{mix}}$ on the detonation front, and a back-propagation shock is formed upstream. In the following cases, $x_{\text{mix}} = 15$ mm is adopted.

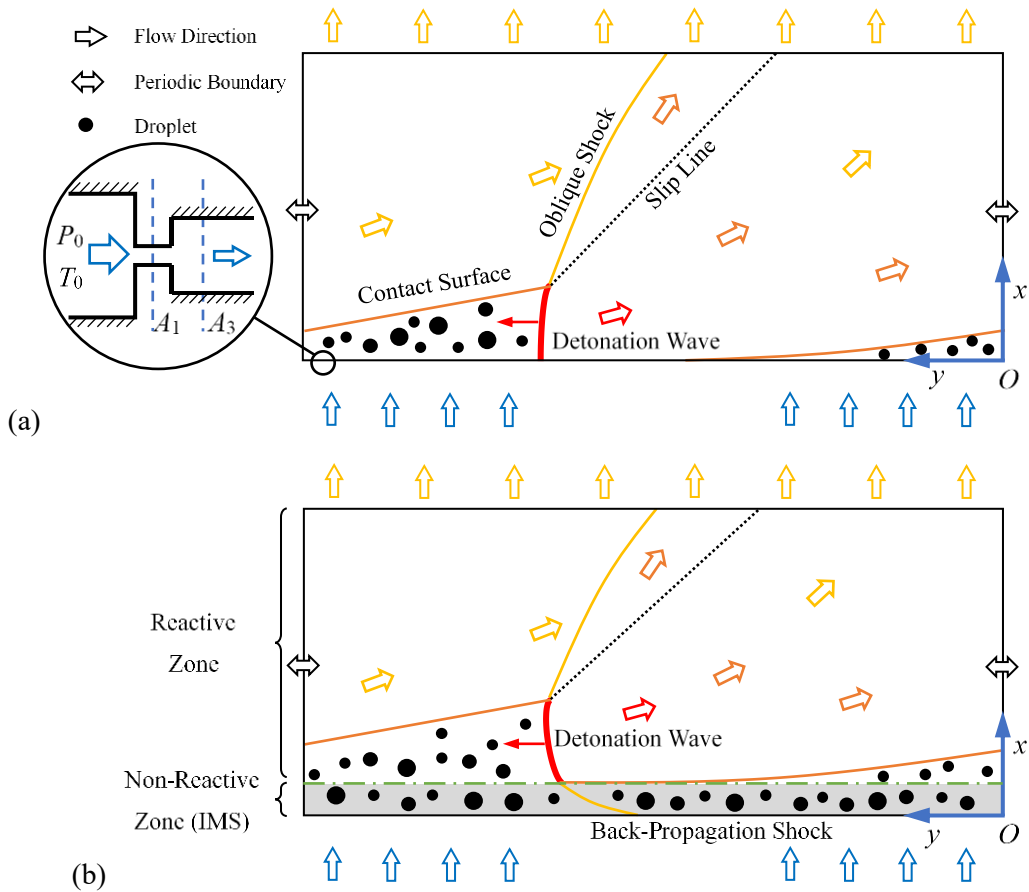

**Figure 1** Schematics of two types of two-dimensional droplet-laden two-phase rotating detonation combustor: (a) Combustor A without IMS; (b) Combustor B with IMS.

The premixed mixture of oxidizer and gaseous kerosene laden with kerosene droplets is injected into the combustor. The total pressure is $P_0 = 700$ kPa and the total temperature is $T_0 = 300$ K for the mixture at the inlet. Define the total spray equivalence ratio as

$$\varphi_t = \frac{\dot{m}_{gk} + \dot{m}_{dk}}{\dot{m}_{go} (F/O)_{st}}$$

and the pre-evaporated equivalence ratio as

$$\varphi_{pre} = \frac{\dot{m}_{gk}}{\dot{m}_{go} (F/O)_{st}}$$

where $\dot{m}_{gk}$ is mass flow rate of gaseous kerosene, $\dot{m}_{dk}$ is the mass flow rate of the fuel droplet, $\dot{m}_{go}$ is the mass flow rate of the oxidizer and $(F/O)_{st} = 0.1261$ is the mass ratio of chemical stoichiometric fuel and oxidizer.

The Rosin-Rammler distribution is adopted to generate kerosene droplets of random size, to fit the droplet size distribution in the experiments. The Rosin-Rammler distribution can be expressed as

$$R(d_p) = 100 e^{-(d_p/d_0)^n} \qquad (23)$$

Here $R(d_p)$ is the cumulative percent passing, which refers to the integral proportion of the volume of all droplets in the range of droplet size $[0, d_p]$. $d_0$ is the average droplet size. When $d_0 = d_p$, $R(d_p) = 36.8\%$. $n$ is the distribution parameter, indicating the concentration of the distribution, and it is fixed as 2 in the study. Too small and too large droplet size generated by the Rosin-Rammler distribution is ignored in the study. The minimum droplet size injected into the combustor is set as $d_p = 0.2 d_0$, and its corresponding $R(d_p)$ is 96.1%. The maximum droplet size is set as $d_p = 2 d_0$, and its corresponding $R(d_p)$ is 1.8%. The initial droplet temperature is 300 K. The effect of initial droplet velocity is not taken into account in the following cases. The fixed velocity $u_d = 50$ m/s and $v_d = v_g$ are adopted, where $v_g$ is the transverse velocity of the local gas flow.

*3.2 Simulation case*

The parameters for all simulation cases shown in the study are listed in Table 1. The two combustors are distinguished by A/B. For example, Case A4-A represents the simulation case for Combustor A with the condition of A4 in Table 1. The completely pre-evaporated cases (namely $\varphi_{pre} = \varphi_t$) and the partially pre-evaporated cases are successively analyzed below.

**Table 1** Parameters for simulation case ($P_0 = 700$ kPa, $T_0 = 300$ K)

| Case | $d_0$ ($\mu$m) | $\varphi_t$ | $\varphi_{pre}$ |
|------|----------------|-------------|-----------------|
| A4   |                | 0.4         | 0.4             |
| A5   | 0              | 0.5         | 0.5             |
| A7   |                | 0.7         | 0.7             |
| B3   | 3              |             |                 |
| B4   | 4              |             |                 |
| B5   | 5              | 0.7         | 0.4             |
| B6   | 6              |             |                 |
| B7   | 7              |             |                 |

**4 Propagation Behaviors of RDW in Completely Pre-Evaporated Cases**

In order to study the influence of droplets on the propagation of RDW, the flow field characteristics in the completely premixed and pre-evaporated cases are firstly analyzed. Additionally, the detonative limit of the studied reactant is numerically obtained.

Firstly, the effects of the IMS on the flow structures are discussed. Figure 2 and Figure 3 show the contours of temperature, pressure and kerosene mass fraction in the Combustor A and B under $\varphi_{pre} = \varphi_t = 0.5$, as well as the detonation cellular structure recorded by the numerical soot-foil technique. The flow field in the Combustor A is similar to the other numerical simulations of gaseous RDW [41,42]. The main structures including the detonation wave, the oblique shock, the contact surface and the slip line can be clearly distinguished by the temperature contour in Figure 2. Multiple transverse waves are propagating on the detonation front to maintain the stable propagation of RDW. The gaseous kerosene is evenly distributed in the refilled zone and reacts completely on the detonation front. The cellular structure is regular and uniform, indicating that the RDW propagates stably. From the partial enlarged view of the cellular structure, the transverse wave propagating downstream penetrates the burned gas at the triple point of detonation wave, and then a weak reflected transverse wave is formed and propagates upstream. At the inlet boundary, the transverse wave interacts with the inlet boundary and a strong reflected transverse wave is formed there.

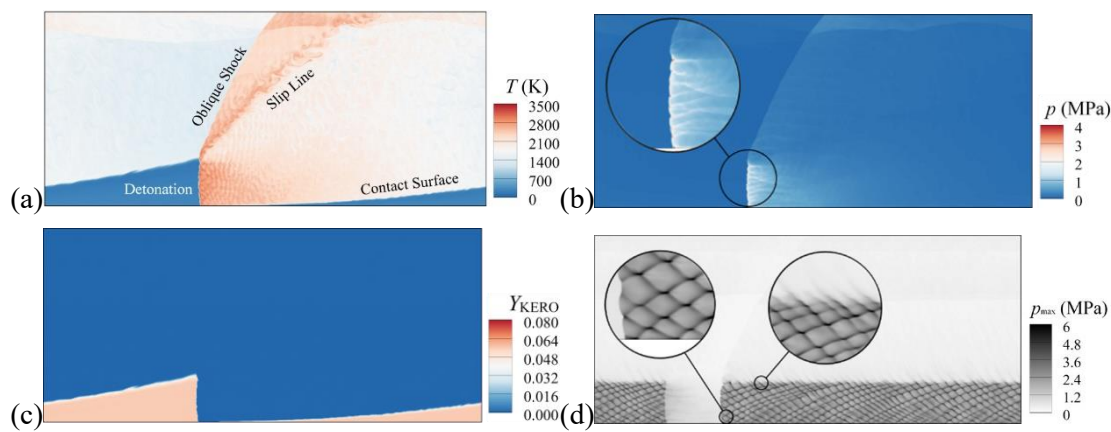

**Figure 2** Contours of (a) temperature, (b) pressure and (c) kerosene mass fraction, and (d) the numerical soot-foil record for Case A5-A ($\varphi_{pre} = \varphi_t = 0.5$).

In the Combustor B, due to the existence of IMS, the detonation wave induces a back-propagation shock in the upstream field, which reflects at the inlet to form a reflected shock, as shown in Figure 3. The back-propagation shock and the reflected shock heat a part of the gas in the refilled zone, so that two layers of reactant with different temperatures are formed in the front of the detonation wave, namely the preheating zone and the non-preheating zone. Figure 4 shows the distribution of static temperature along $y$-direction in front of detonation wave. Due to the acceleration of the gas flow, the static temperature gradually drops in the refilled zone of the two combustors. The static temperature in Combustor A drops from 286 K to 180 K at $x =$ 40 mm. In the Combustor B, the temperature drops to 193 K at $x =$ 30 mm and then rises to about 350 K in the preheating zone.

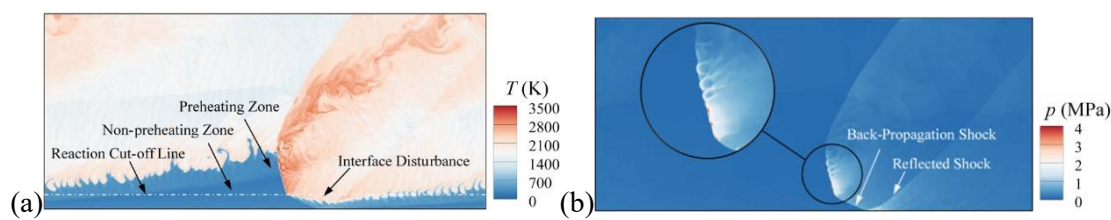

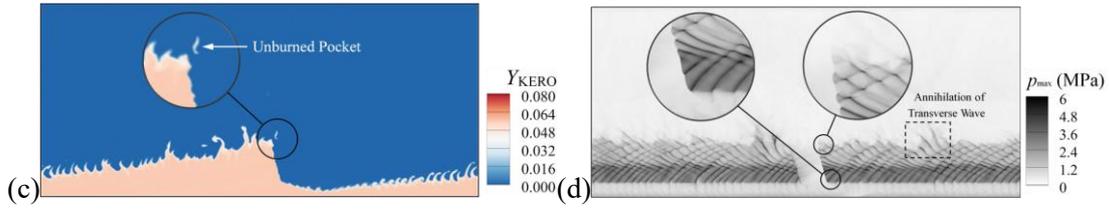

**Figure 3** Contours of (a) temperature, (b) pressure and (c) kerosene mass fraction, and (d) the numerical soot-foil record for Case A5-B ($\varphi_{pre} = \varphi_t = 0.5$).

Due to the layered temperature, the intensity of transverse waves and the cellular structure also form two layers (see Figure 3). In the preheating zone, because of the higher temperature and lower density, the detonation peak pressure is significantly reduced compared with the non-preheating zone, and the cell size is smaller. Unlike in the Combustor A, there is no wall restriction on the upstream of the detonation wave in the Combustor B. The transverse waves penetrate the gaseous layers at both the upstream and downstream triple points. However, due to the difference in temperature and composition of the two layers, the acoustic impedance of the gaseous layers is different, and thus the behavior of the transmitted transverse waves is also different [43]. At the downstream triple point, similar to the Combustor A, a new reflected transverse wave is formed. While at the upstream triple point, the transverse waves attenuate after transmission and no reflected transverse wave is formed. The transverse waves induce interface disturbance of the contact surface at the upstream triple point. The interface disturbance increases in a limited amplitude as it develops downstream. The temperature and composition disturbance on the contact surface results in the disappearance of transverse waves at some locations, which adversely affects the stability of the detonation. Additionally, the disturbance also introduces some small unburned pockets after the detonation wave and they are burned downstream in the deflagration mode.

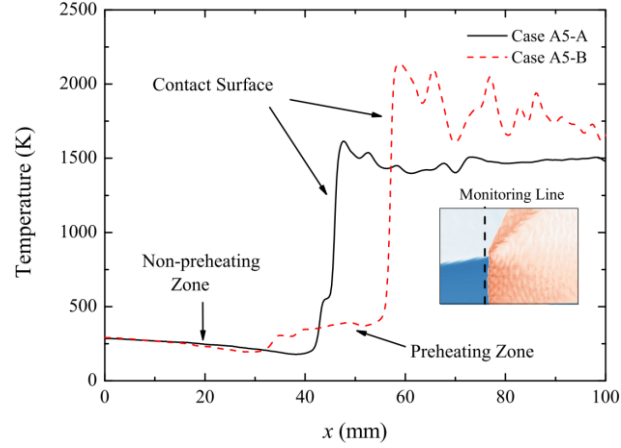

**Figure 4** Distribution of static temperature along *y*-direction in front of detonation wave.

The RDW can propagate stably at higher equivalence ratio $\varphi_{pre} = \varphi_t = 0.7$. Figure 5 shows the temperature contour and numerical soot-foil record of Case A7. The main structures of the flow field are consistent with those of Case A5, while the transverse waves are less obvious and the cell size is much smaller, and thus the interface disturbance is significantly weakened.

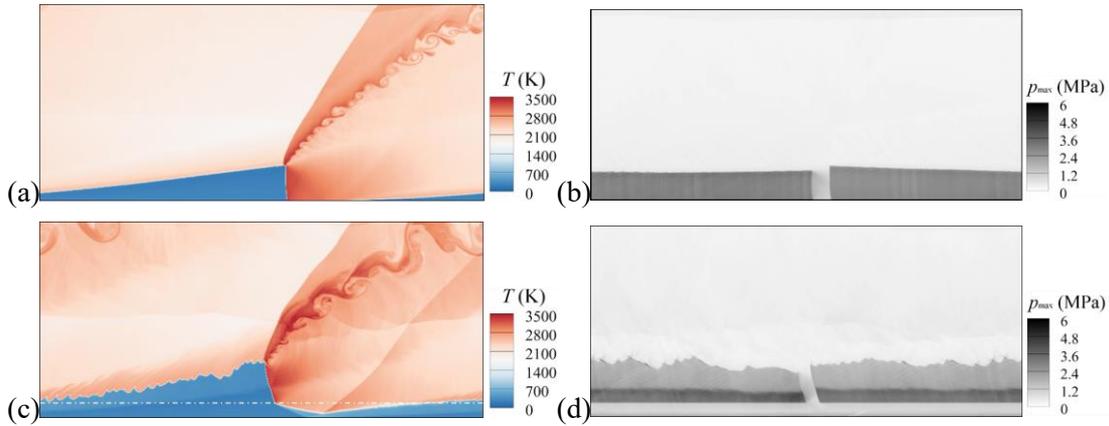

**Figure 5** Temperature contour and numerical soot-foil record of (a)-(b) Case A7-A and (c)-(d) Case A7-B ($\varphi_{pre} = \varphi_t = 0.7$).

The temperature variation in the preheating zone in the Combustor B is further analyzed in Figure 6. Obviously, the preheating zone can be divided into three zones with different temperatures. Zone (I) and Zone (II) are respectively the heating zone after the back-propagation shock and the reflected shock. Zone (III) is the expansion zone after the reflected shock and it extends to the detonation front. The temperature continuously decreases in Zone (III). In Case A7-B, the average temperatures in the three zones are approximately 700 K, 640 K and 340 K, respectively. Due to the disturbance in Zone I by transverse waves, the intensity of the back-propagation shock in Case A5-B is weakened and the three temperatures drops to 590 K, 625 K, 320 K, respectively.

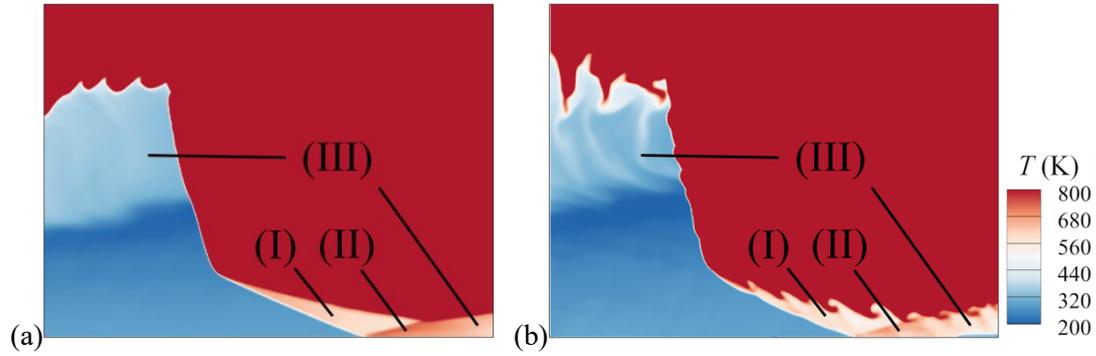

**Figure 6** Temperature contour in the vicinity of combustion inlet for (a) Case A7-B and (b) Case A5-B. Zone (I): heating zone after the back-propagation shock. Zone (II): heating zone after the reflected shock. Zone (III): expansion zone after the reflected shock.

When the total equivalence ratio $\varphi_t$ is lower than 0.5, the RDW cannot sustain. The detonation quenching process when $\varphi_t$ is reduced from 0.5 to 0.4 is shown in Figure 7 and Figure 8. The characteristics of the detonation quenching process in the Combustor A and Combustor B are different. As shown in Figure 7a, a few of unburned pockets are generated at the triple point in the Combustor A. The transverse waves cannot sustain and continuously attenuate. Finally, the incident shock and the reaction zone are totally decoupled, and the detonation quenches. Correspondingly, the cell size gradually increases during the quenching process, as shown in Figure 8a. In the Combustor B, the decoupling first occurs in the vicinity of the upstream triple point, where the transverse waves annihilate due to the low equivalence ratio. The decoupling zone gradually expands downstream and finally leads to the detonation quenching.

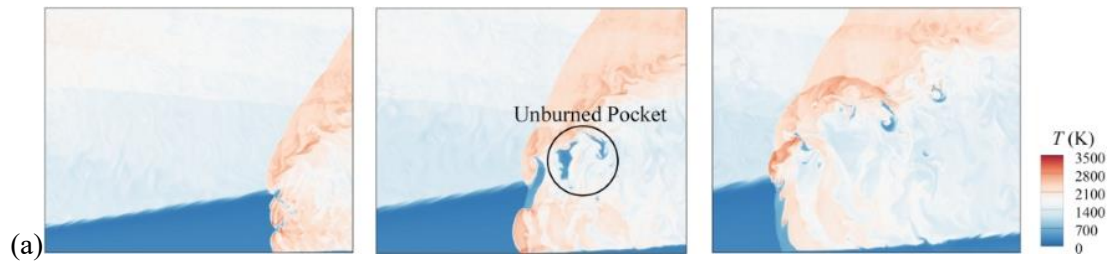

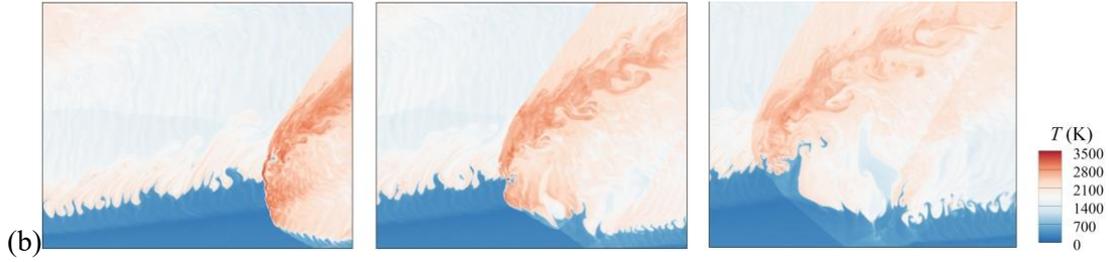

(b)

**Figure 7** Time sequences of temperature contour during the detonation quenching process in (a) Case A4-A and (b) Case A4-B ($\varphi_{pre} = \varphi_t = 0.4$). The time interval is 50 $\mu$s.

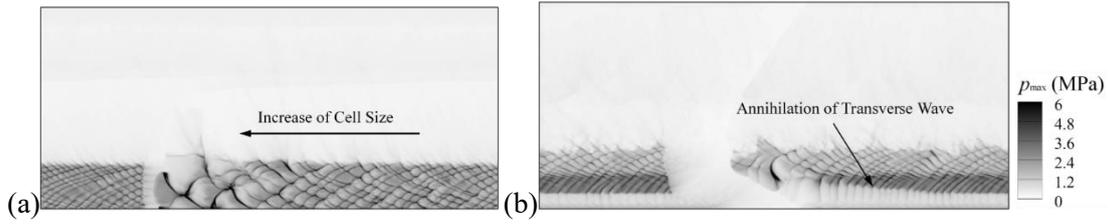

(a)       (b)

**Figure 8** Numerical soot-foil record shown the detonation quenching process in (a) Case A4-A and (b) Case A4-B ($\varphi_{pre} = \varphi_t = 0.4$).

Furthermore, the detonation wave velocities under different equivalence ratio are compared in Figure 9. Due to the lack of experimental data of the kerosene/oxygen-enriched air detonation, the theoretical Chapman-Jouguet velocity $V_{CJ}$ of Jet-A/oxygen-enriched air detonation is calculated using CEA software [44] as a reference. Because the fuel is completely pre-evaporated and premixed, the local equivalence ratios in the refilled zone are uniform and the same in the two combustors. Therefore, the detonation wave velocities $V_D$ are also approximately the same, and close to $V_{CJ}$. Based on the simulation results, the lower detonative limit of the studied reactant is $\varphi_{lb} = 0.5$.

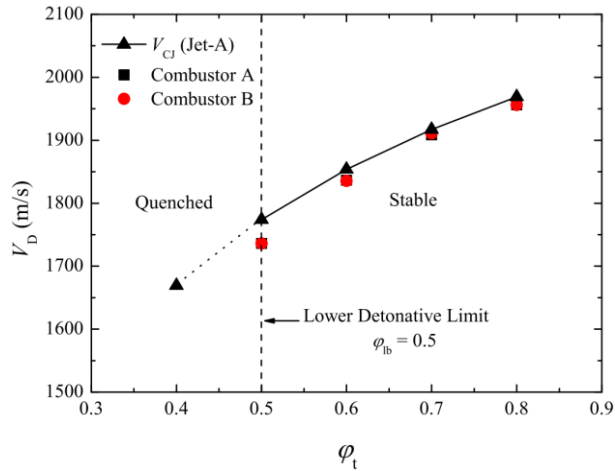

**Figure 9** The comparison of detonation wave velocity under different $\varphi_t$ for the completely

premixed and pre-evaporated cases.

**5 Droplet Evaporation Mechanisms in Partially Pre-evaporated Cases**

Given $\varphi_t = 0.7$, $\varphi_{pre} = 0.4$ and $d_0 = 3$ $\mu$m, the two types of droplet evaporation mechanism in Combustor A and B are analyze and compared in this section.

Although the inlet pre-evaporation equivalence ratio $\varphi_{pre} = 0.4$ is less than $\varphi_{lb}$, the RDW can propagate stably in the two combustors, as shown in Figure 10 and Figure 11. While, the propagation behavior and stability mechanism are different. In the Combustor A, the kerosene mass fraction in the refilled zone does not change because the droplets hardly evaporate. As shown in the droplet distribution diagram (Figure 10e), the majority of droplets rapidly evaporate and the reaction is completed within a short distance after the detonation wave, thereby promoting the stable propagation of the detonation wave. Since the droplet evaporation and reaction is gradually completed, the high temperature area after the detonation wave is larger compared to Case A7-A. Because the airflow in the refilled zone continuously accelerates in the $x$-direction, there is always a speed difference between the droplet and the local airflow. Therefore, a droplet-free band is formed in the vicinity of contact surface (see Figure 10e), where the local equivalence ratio is below $\varphi_{lb}$, and the incident shock and reaction zone are locally decoupled. The transverse wave is annihilated when propagating there. A few of unburned pockets are generated at the triple point and consumed downstream. From Figure 10d, the cellular structure is more irregular and the cell size is bigger compared with the completely pre-evaporated case. In addition, the transverse wave annihilation can be observed more frequently, indicating the detonation propagation is less stable.

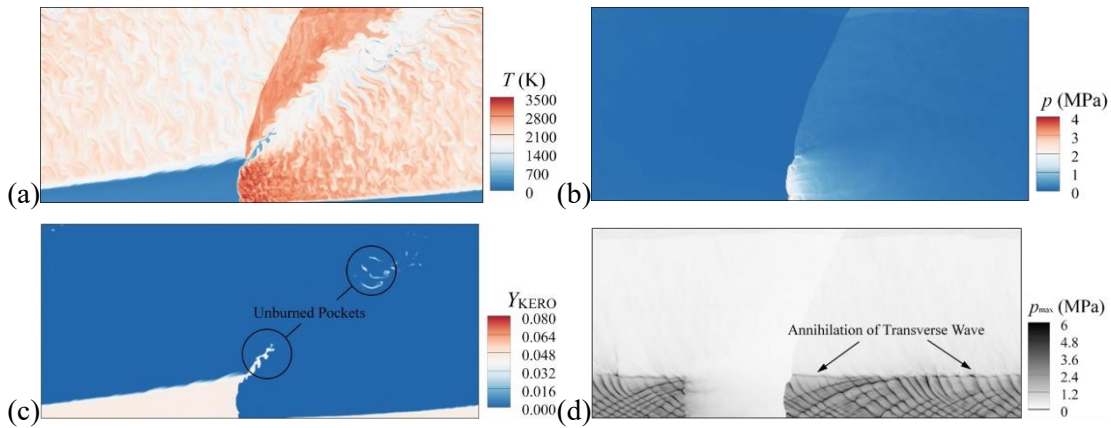

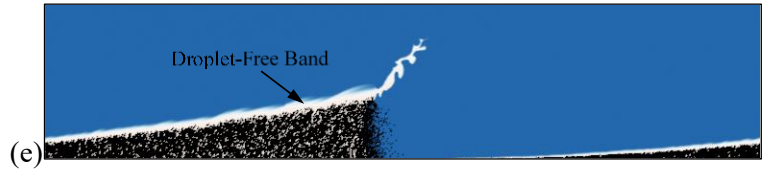
(e)

**Figure 10** Contours of (a) temperature, (b) pressure and (c) kerosene mass fraction, and (d) numerical soot-foil record, (e) distribution of droplets for Case B3-A ($\varphi_t = 0.7$, $\varphi_{pre} = 0.4$ and $d_0 = 3\ \mu m$). Each black point denotes a droplet.

As shown in Figure 11e, the droplet distribution in the Combustor B forms two obvious layers. A large number of droplets gather in the non-preheating zone, while in the pre-heating zone, the droplets continuously evaporate, and the number of droplets is continuously reduced. In front of the detonation, the majority of droplets have completely evaporated. The rest of droplets with large initial diameters evaporate after the detonation front. Correspondingly, the distribution of gaseous kerosene is also stratified (Figure 11c). The equivalence ratio in the non-preheating zone is approximately equal to $\varphi_{pre} = 0.4$, while it is close to $\varphi_t = 0.7$ in the preheating zone. In addition, the disturbance of the contact surface is evidently strengthened compared to Case A5-B. The interface is greatly stretched and deformed in some locations, thereby introducing a few of unburned pockets into the combustion production. The producing region and mechanism of the unburned pockets are different from those in the Combustor A. This disturbance propagates back to the refilled zone and affects the droplets distribution, resulting in the non-uniform distribution of kerosene mass fraction in the preheating zone. Since the local equivalence ratio in the preheating zone is high, the transverse waves on the detonation front are very weak. The cellular structure is not easy to be distinguished from the numerical soot-foil record (Figure 11d). The interface disturbance causes the periodic stretching and compression of the detonation front, and forms the oscillation in cellular structure. The strong transverse waves are observed during the stretching process.

It is worth noting that due to the layered composition, the equivalence ratio in the non-preheating zone is too low to maintain the detonation propagation. Therefore, different from Case A5-B and Case A7-B, the RDW can self-sustain away from the given reaction cut-off line $y = 15$ mm in Case B3-B (Figure 11a).

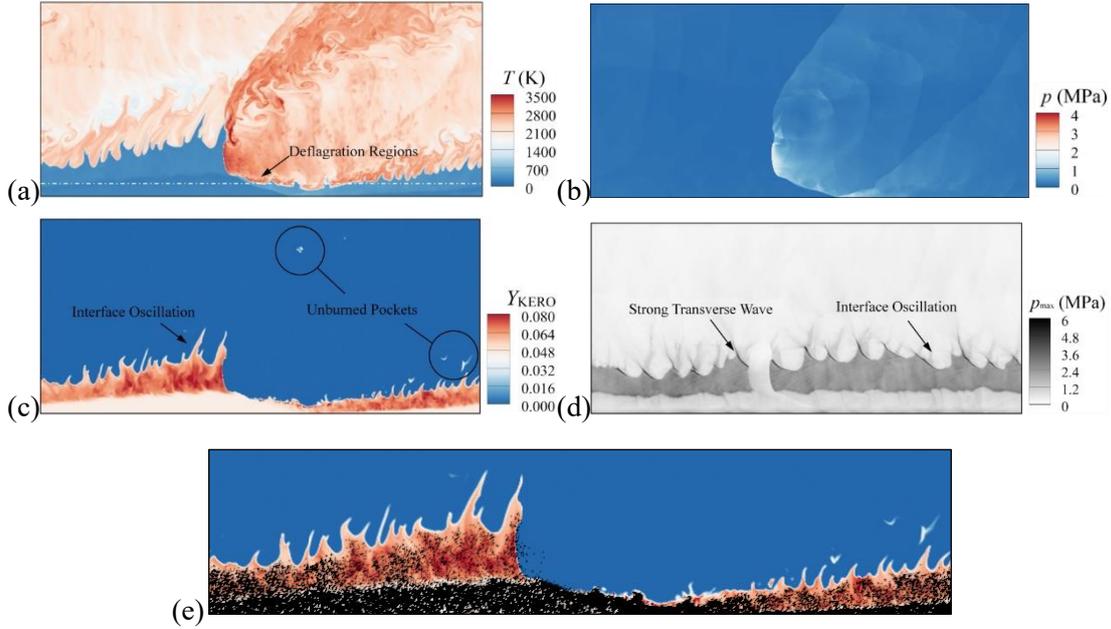

**Figure 11** Contours of (a) temperature, (b) pressure and (c) kerosene mass fraction, and (d) numerical soot-foil record, (e) distribution of droplets for Case B3-B ($\varphi_t = 0.7$, $\varphi_{pre} = 0.4$ and $d_0 = 3~\mu$m). Each black point denotes a droplet.

Figure 12 and 13 respectively show the distribution of the intial size and the cumulative evaporation mass of droplets in the vicinity of the detonaiton wave. The cumulative evaporation mass is defined as the difference between the intial droplet mass and the current mass. In these figures, each black circle denotes a droplet and the circle diameter represents the relative value of the studied variables. In the Combustor A, droplets of different initial diameters are evenly distributed in the refilled zone. The small droplets are more likely to gathered near the contact surface because of the the small inertia and good followability, as shown in Figure 12a. The droplet evaporation mainly completes in a short distance of 15 mm after the detonation front (Figure 13a). In the Combustor B, droplets gather in the non-preheating zone after entering the combustor. Shortly, the droplets encounter with the back-propagation shock and are carried upstream into the pre-heating zone. From Figure 13b, the cumulative evaporation mass of droplets increases signficantly in the pre-heating zone. After the back-propagation shock, a few of large droplets pass through the contact surface and evaporate quickly in the high-temperature environment to form the deflagration regions shown in Figure 11a and Figure 13b. The small droplets can evaporate rapidly in the preheating zone. Only a small amount of large droplets remain in front of the detonation wave and eventually evaporate after the detonation.

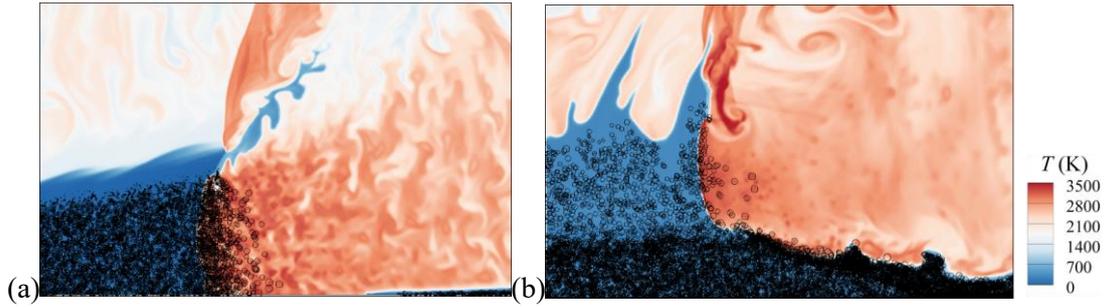

**Figure 12** Distribution of initial droplet diameter in the vicinity of detonation wave in (a) Combustor A and (b) Combustor B. Each black circle denotes a droplet and the circle diameter represents the relative value of the initial diameter.

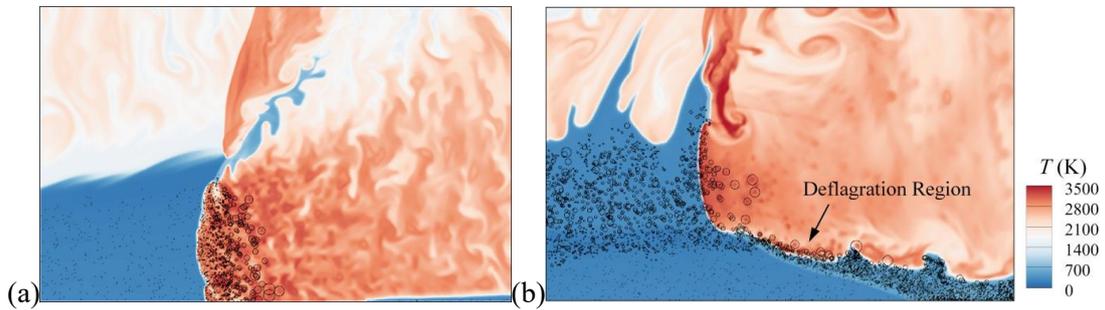

**Figure 13** Distribution of cumulative evaporation mass in the vicinity of detonation wave in (a) Combustor A and (b) Combustor B. Each black circle denotes a droplet and the circle diameter represents the relative value of the cumulative evaporation mass.

In order to further understand the dynamics characteristics and evaporation process of droplets, three droplets with an initial diameter $d_0 = 3$ $\mu$m are released and tracked at the inlet boundary, as shown in Figure 14. The interval between the droplets is the same and Particle A is closest to the detonation front or back-propagation shock. The trajectories of the tracked droplets are shown in Figure 15, where the thickness of the line represents the droplet diameter. The variations of the droplet diameter, temperature and velocity are compared in Figure 16.

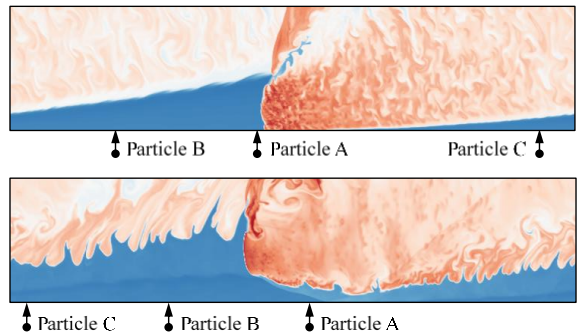

**Figure 14** Schematics of release position of the tracked droplets in two combustors.

The dynamics characteristics of the three droplets in the two combustors are basically similar. The detonation wave first encounters Particle A and carries the droplet to the positive *y*-direction. Shortly, the droplet completely evaporates in 5 $\mu$s. As can be seen in Figure 16b, the temperature of Particle A first rises to 600 K in the induction zone after the incident shock, that is the Von Neuman state of the detonation. Then the droplet is heated again in the high-temperature reaction zone, and evaporates quickly. Because the transverse waves propagating in the opposite direction induce the opposite *x*-velocity after the detonation, the velocity of Particle A also decreases first and increases again and shows a U-shape variation (Figure 16c). Before Particle B encounters the detonation wave, it gradually accelerates and cools down in the refilled zone. Its behavior after the detonation is similar to that of Particle A. Particle C is in contact with the detonation at *x* = 21 mm, which is not shown in Figure 15a.

The movement path of droplets in Combustor B is more complicated. Particle A encounters the back-propagation shock shortly after entering the combustor, and is pushed upstream by the back-propagation shock (Figure 15b). Because Particle A is close to the inlet boundary, it encounters the reflected shock soon. In the preheating zone, the droplet temperature quickly rises to 450 K and reaches equilibrium with the ambient temperature (Figure 16b). Meanwhile, the droplet evaporates and the droplet size decreases. However, Particle A leaves the preheating zone near the inlet boundary at 40 $\mu$s and enters the non-preheating zone again. At this time, the droplet size has been reduced from 3 $\mu$m to 1.2 $\mu$m (Figure 16a). Then, Particle A gradually accelerates and moves downstream, while its temperature drops below 300 K and the evaporation stops. Before encountering the back-propagation shock in the next cycle, Particle A remain in the non-preheating zone.

From Figure 16c, Particle B continuously accelerates to $u_d$ = 150 m/s after entering Combustor B, and has moved to *x* = 7 mm before encountering the back-propagation shock. The residence time of Particle B in the preheating zone is longer than that of Particle A. After encountering the back-propagation shock, the velocity of Particle B $u_d$ decreases to -300 m/s, and its trajectory also deflects to the upstream (Figure 15b), while the droplet temperature rises to about 410 K (Figure 16b). Soon, Particle B encounters the reflected shock. Its velocity increases to 100 m/s and Particle B moves downstream again. The temperature rises again to 470 K and the evaporation is completed.

The behaviors of Particle C and B are similar, expect that the contact position of Particle C with the back-propagation shock is more downstream. Since the contact position with the back-propagation shock of three droplets is different, the residence time in the preheating zone is also different. Based on the droplet velocity change in Figure 16c, the residence time for three droplets are calculated as $\Delta t_A = 5\ \mu s$, $\Delta t_B = 20\ \mu s$ and $\Delta t_C = 30\ \mu s$.

Note that, since the tracked droplets are small and completely evaporate in the preheating zone, they do not encounter the detonation wave during the entire life. However, for the droplets of larger diameter, the evaporation time is significantly prolonged. Those droplets whose release position is close to the back-propagation shock can re-enter the preheating zone for several times and evaporate, similar with Particle A. While the droplets far away from the back-propagation finally evaporate after the detonation wave as aforementioned in Figure 12.

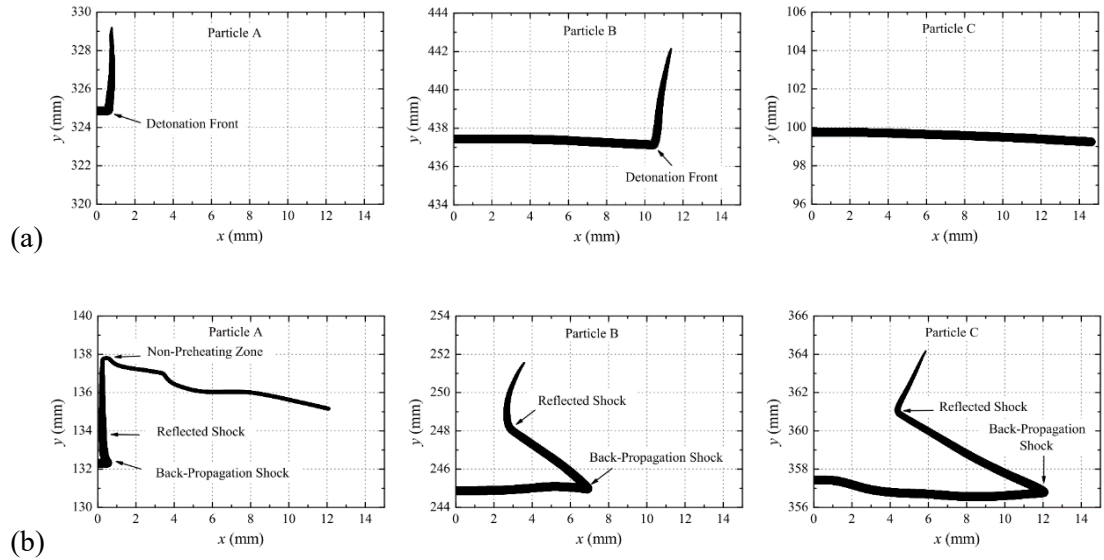

**Figure** 15 Trajectories of the tracked droplets in (a) Combustor A and (b) Combustor B. The initial diameter of the droplet is 3 $\mu$m. The scale of the $x$ and $y$ coordinates in the figure is equal.

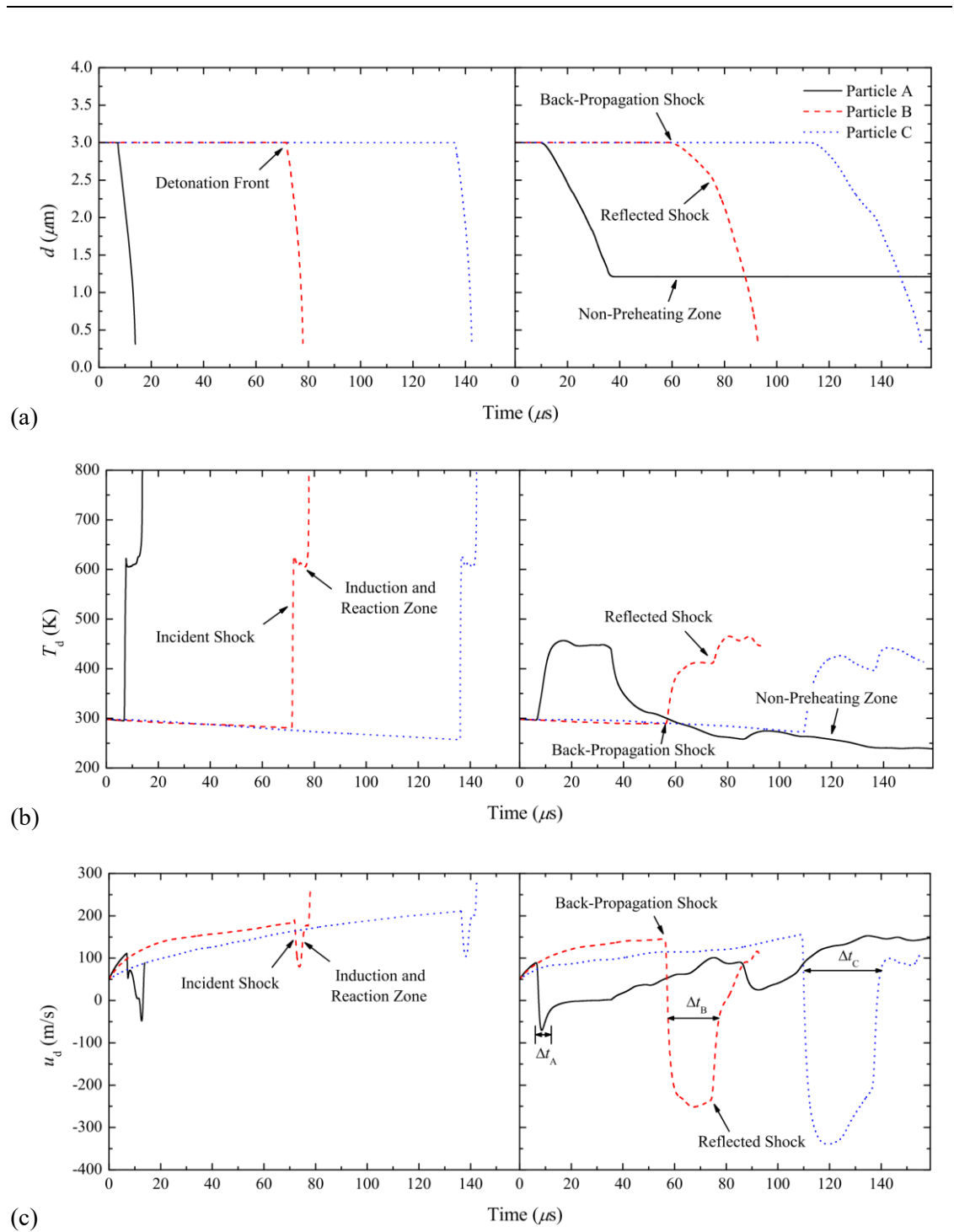

**Figure 16** Variation of the droplet parameters with time. (a) Droplet diameter $d$; (b) Droplet temperature $T_d$ and (c) Droplet velocity component in $x$-direction $u_d$. The left column shows the parameters in Combustor A, and the right column in Combustor B.

Figure 17 shows the pressure histories recorded at the monitoring points in Combustor A and B. The monitoring points are respectively located at $(x, y) = (1, 0)$ mm and $(26, 0)$ mm, close to the upstream boundary of detonation front. The RDW propagates stably for 7 cycles in the shown time. The detonation pressure peak fluctuates between 2.0 and 3.5 MPa during these cycles. Because the density in the preheating zone is reduced, the pressure peak in the Combustor B is slightly smaller than that in the Combustor A. A small peak can be identified on the pressure signal in the Combustor B, which corresponds to the reflected shock.

The average detonation wave velocity can be calculated by $V_D = L_y/\Delta t_D$, where $L_y$ =0.45 m is the $y$-direction dimension of the combustor, $\Delta t_D$ is the average period of the RDW during seven cycles. Then, the detonation wave velocities of the two combustors are $V_{D,A}$ = 1,713 m/s and $V_{D,B}$ = 1,874 m/s, respectively. The theoretical velocity for $\varphi_t$ = 0.7 is $V_{CJ}$ = 1,915 m/s, and thus $V_{D,A}$ = 0.90 $V_{CJ}$, $V_{D,B}$ = 0.98 $V_{CJ}$. The detonation velocity deficit of Combustor B is obviously smaller than that of Combustor A. Note that, $V_{D,A}$ is close to the wave velocity of Case A5-A ($\varphi_t = \varphi_{pre}$ = 0.5) that is equal to 1735 m/s. This indicates that the droplet heat release after the detonation wave is partially coupled with the incident shock, and promotes the detonation propagation. Obviously, the initial droplet diameter will affect the coupling degree of heat releasing, and then affect the propagation stability of the RDW, which will be further analyzed in the next section.

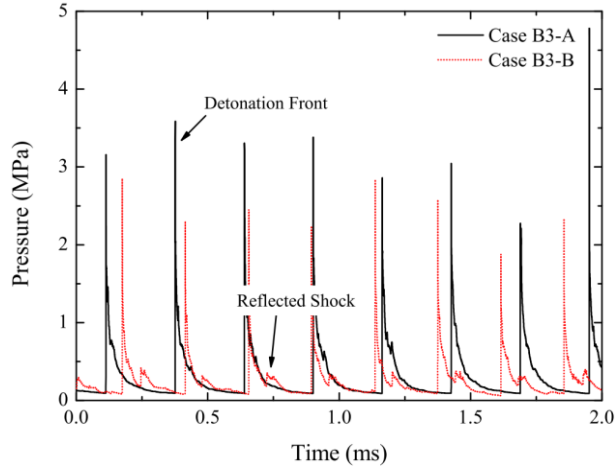

**Figure 17** Pressure histories at the monitoring points in the combustor for Case B3.

**6 Sensitivity of Detonation Propagation Stability to the Average Droplet Diameter**

The sensitivity of detonation propagation stability to the average droplet size is studied in this section, and the instability mechanism of RDW is analyzed. The simulation cases shown in this section are based on Case B3, that is increasing the average droplet diameter $d_0$ gradually and keeping the other parameters unchanged (namely $\varphi_t$ = 0.7, $\varphi_{pre}$ = 0.4).

In the Combustor A, the evaporation distance after the detonation wave increases significantly with the increase of $d_0$, and thus the coupling between the incident shock and reaction zone is weakened, and the detonation tends to unstable. When $d_0$ increased from 3 to 4 $\mu$m, the RDW comes to a critically stable state. Figure 18 shows the cellular structure during 0.5 ms, as well as the contours of temperature and fuel mass fraction at several critical moments. The cumulative evaporation mass is represented by the circle size in the temperature contour. The RDW goes through several stages including stable propagation, quenching, re-ignition and re-stabilization. Before time instant I, the RDW has propagated stably for several cycles. The droplet evaporation distance after the detonation wave is longer when $d_0$ increases to 4 $\mu$m. Similar to Case B3-A, the annihilation of transverse waves can be observed frequently during the detonation propagation. In most cases, new transverse waves will be formed from the detonation triple point to maintain the stable propagation. However, after the transverse wave is annihilated at time instant II, the new transverse wave is not formed immediately and thus the cellular structure is broken. At time instant III, the incident shock is decoupled with the reaction zone, and the detonation wave quenches. In the high temperature zone after the incident shock wave, a large number of droplets evaporate quickly, and a mixing region of high equivalence ratio is formed in front of the reaction zone. At time instant IV, the mixing reactant is ignited by the high temperature and a strong transverse detonation is formed. Soon, an overdriven detonation is re-ignited and the RDW continues to propagate int the subsequent cycles. When $d_0$ increases to 4 $\mu$m, the cellular structure is more irregular, and the re-ignition process is observed. Therefore, this state is defined as a critically stable state.

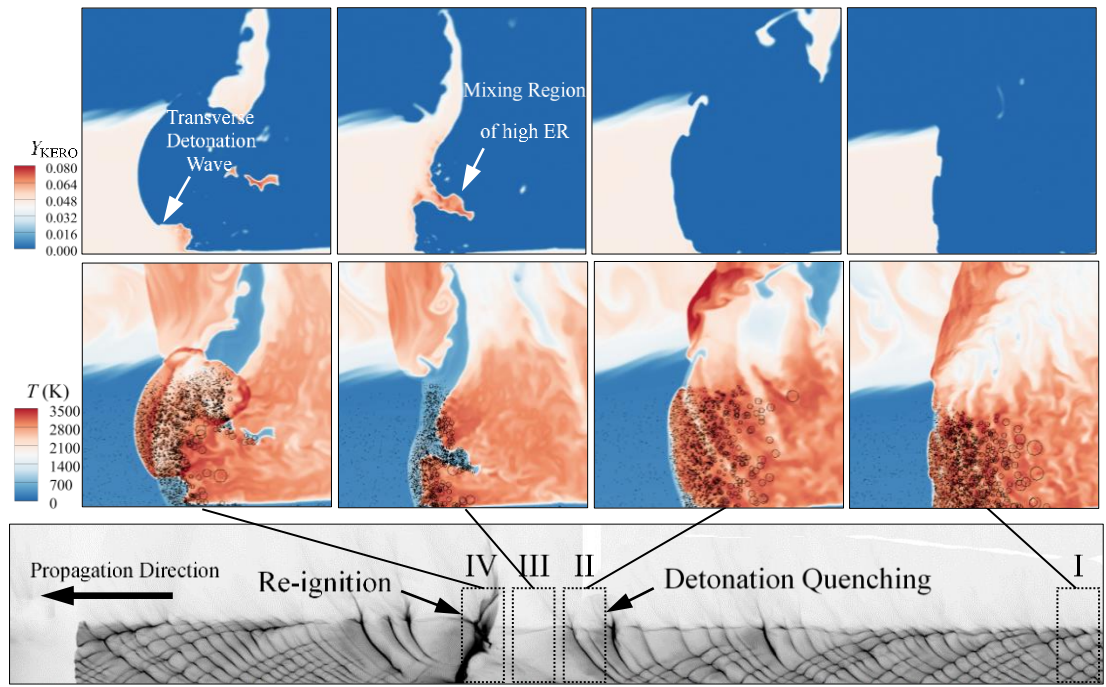

**Figure 18** The process of detonation quenching and re-ignition in Case B4-A. Each black circle denotes a droplet and the circle diameter represents the relative value of the cumulative evaporation mass.

Figure 19 shows the pressure history at the inlet boundary in Case B4-A. The average detonation velocity during the stable propagation period is about 1,675 m/s, and the peak pressure is about 2.5 MPa. Detonation quenching and re-ignition process occurs within 1.0~1.2 ms, and the velocity drops to 1,562 m/s. In the subsequent period, the detonation wave resumes stable propagation, and the velocity rises again.

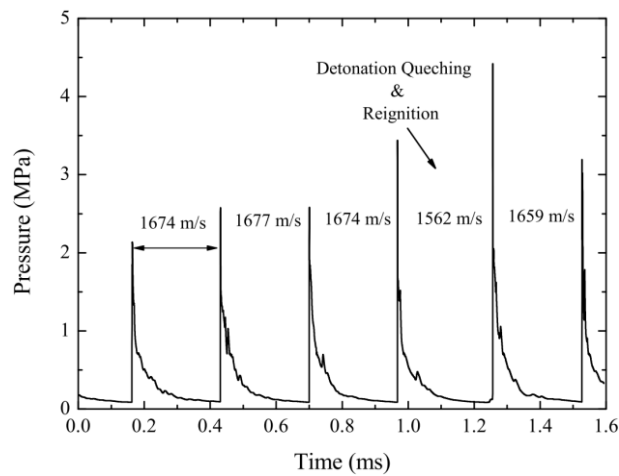

**Figure 19** Pressure history at the inlet monitoring point in Case B4-A. The average detonation velocity during each cycle is shown.

When $d_0$ increases to 5 μm, the detonation wave quenches after a short time. This process is shown in Figure 20. The time interval between the subfigures is 25 $\mu$s. Obviously, the droplet evaporation distance continuously increases for the larger $d_0$. Similar with Case B3-A, an unburned pocket is introduced by the droplet-free band. Due to the large size of droplets, the evaporation mass is insufficient in the induction zone and cannot support the transverse wave. Thus, the unburned region gradually enlarges and the incident shock is fully decoupled with reaction zone at 50 $\mu$s. As shown in the last kerosene mass fraction contour, the droplets evaporate in the induction zone, just like Case B4-A. However, because of the insufficient evaporation, the re-ignition does not occur and the detonation wave quenches in the end.

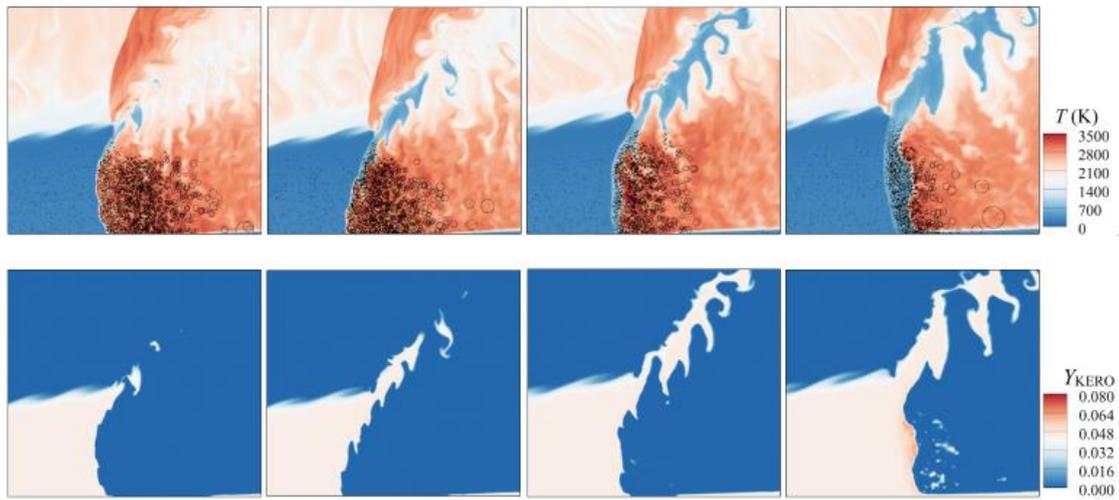

**Figure 20** Time sequences showing the detonation quenching process in Case B5-A. The time interval is 25 $\mu$s. Each black circle denotes a droplet and the circle diameter represents the relative value of the cumulative evaporation mass.

The propagation stability of RDW in the Combustor B is less sensitive to droplet sizes. The droplet evaporation, kerosene mass fraction contour and cellular structure at $d_0$ = 3, 4, 5, and 6 $\mu$m are compared in Figure 21. The RDW can sustain stable propagation for these four droplet sizes. As the droplet size increases, the evaporation of the droplets in the preheating zone is reduced, while more droplets are burned after the detonation. Therefore, the equivalence ratio decreases in front of the detonation wave, and thus the transverse waves are intensified as shown in Figure 21c.

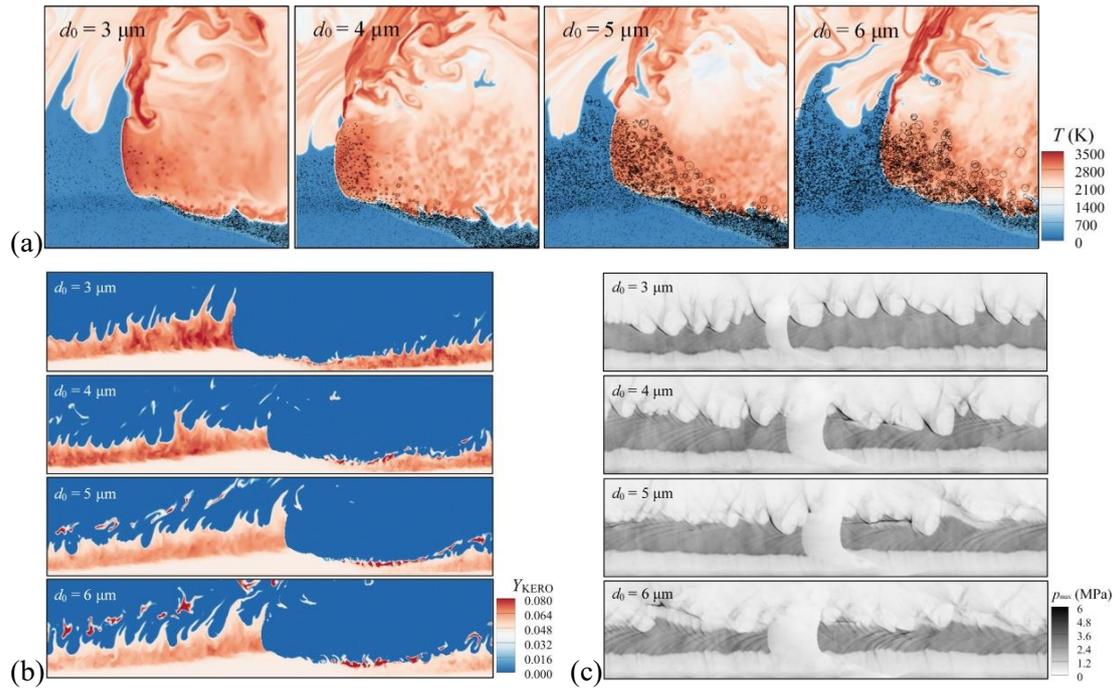

**Figure 21** Comparison of flow fields in the Combustor B for different average droplet diameters. (a) Temperature contour and the cumulative evaporation mass; (b) kerosene mass fraction contour; and (c) numerical soot-foil record.

Figure 22 shows the average equivalence ratio of a 2×2 mm square area in front of the detonation for different $d_0$, where the error bar indicates the variance of equivalence ratio. As $d_0$ increases, the local evquivalence ratio gradually decreases and is lower than the lower detonation limit $\varphi_{lb}$ when $d_0 = 7$ $\mu$m, and the RDW cannot self-sustain. The flow field before the quenching is shown in Figure 23. When $d_0$ changes from 3 $\mu$m to 7 $\mu$m, the local decoupling starts from the upstream boundary of the detonation front, and gradually expands. Finally, the detonation quenches.

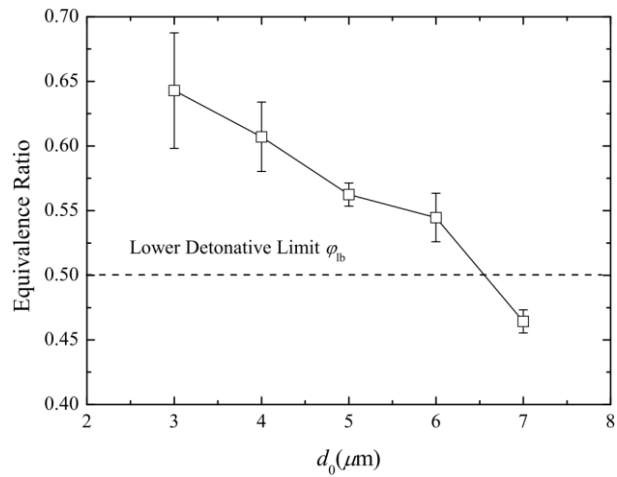

**Figure 22** Average equivalence ratio in front of detoation wave in the Combustor B as a function of $d_0$.

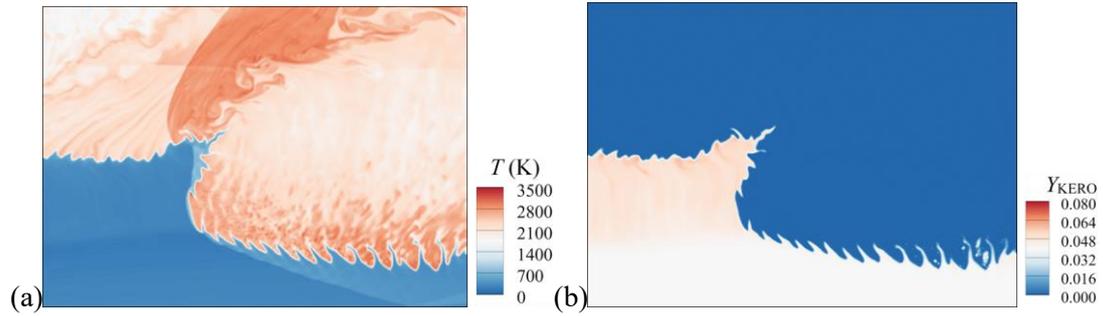

**Figure 23** Contours of (a) temperature and (b) kerosene mass fraction before detonation quenching in Case B7-B.

Finally, the average detonation velocity during five cycles under different $d_0$ is shown in Figure 24, and the status of the RDW is marked. The velocity deficit in the two combustors increases with the increase of the droplet size. The velocity deficit of Combustor B is generally smaller than that of the combustor A. In addition, the maximum droplet size of Combustor B for the stable propagation of detonation wave is also larger. These characteristics indicate that the combustion organization method in Combustor B is more conducive to achieving a stable two-phase RDW.

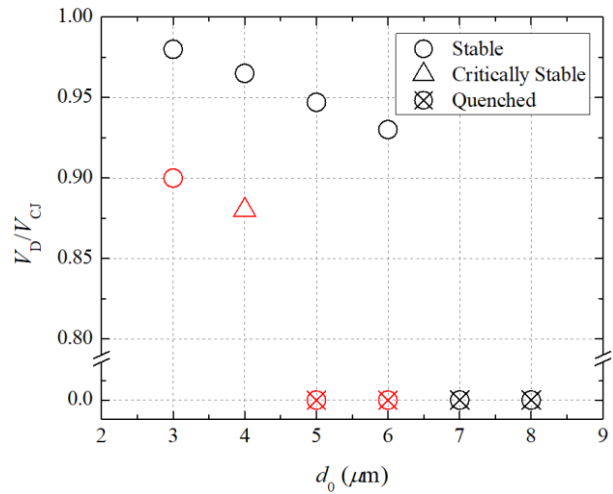

**Figure 24** Variation of detonation wave velocity as a function of average droplet diameter. The combustion status is denoated by red symbols for Combustor A and black symbols for Combustor B.

## 7 Conclusion

Based on the Eulerian-Lagrangian method, two schemes of normal-temperature two-phase rotating detonation system without and with the inlet mixing section (denoted by Combustor A and B respectively) are numerically studied. The oxidizer used in the study is oxygen-enriched air, and the fuel is partially pre-evaporated liquid kerosene. The flow field structure, droplet evaporation mechanism and stabilization mechanism of RDW in the two combustors are compared. The effects of average droplet diameter $d_0$ on the propagation characteristics and instabilities are also analyzed. The main conclusions obtained in the study are as follows.

For the completely pre-evaporated cases in the Combustor B, the detonation wave induces a back-propagation shock and a reflected shock at the combustor head, which preheat the fresh premixed gas in the refilled zone and produce a preheating zone upstream of the contact surface. The temperature in the preheating zone near the back-propagation shock can reach above 600 K, and the temperature gradually drops to about 350 K downstream. The transverse waves at the detonation front stimulate the interface oscillation of the contact surface. When the total spray equivalence ratio $\varphi_t$ is reduced to less than 0.5, the RDW in the two combustors cannot self-sustain and quenches in the end.

For the partially pre-evaporated cases with $\varphi_t = 0.7$ and $\varphi_{pre} = 0.4$, the droplet evaporation mechanism in the two types of combustor is significantly different. In the Combustor A, the droplets do not evaporate in the refilled zone due to the low static temperature. All the droplets evaporate in the induction zone and the reaction zone after the detonation wave. Although the local equivalence ratio in front of the detonation wave is below the lower detonative limit $\varphi_{lb}$, the detonation can maintain stable propagation because the droplets evaporate and complete heat release rapidly in a short distance after the detonation wave. While the detonation cellular structure becomes more irregular under this condition. In the Combustor B, the droplets first enter the non-preheating refilled zone, and enter the preheating zone after encountering with the back-propagation shock, where most of the small-size droplets evaporate completely. The rest of large-size droplets continue to evaporate after the detonation wave. Due to the evaporation in the preheating zone, the local equivalence ratio in front of the detonation wave is approximately equal to $\varphi_t$, and thus the RDW can propagate stably.

As the average droplet diameter $d_0$ increases, the droplet evaporation distance after the detonation becomes larger in Combustor A. Therefore, the coupling between the reaction and the incident shock weakens, and the detonation propagation tends to be less stable. Given the studied reactant, when $d_0$ increases to 4 $\mu$m, the RDW is in a critically stable state. The phenomenon of detonation quenching and re-ignition is observed. When $d_0$ continues to increase to 5 $\mu$m, transverse waves on the detonation front gradually disappear and the incident shock is decoupled with the heat release region, finally resulting in the detonation quenching. In the Combustor B, the droplet evaporation decreases in the preheating zone as $d_0$ increases, and thus the local equivalence ratio in front of the detonation wave is reduced. The transverse waves strengthen and the detonation velocity deficit increases. When $d_0$ increases to 7 $\mu$m, the local equivalence ratio is lower than $\varphi_{lb}$, and the detonation quenches.

The study shows that the combustor with IMS can increase the droplet evaporation in front of the detonation wave, and is more conducive to the propagation stability. It should be noted that the evaporation mechanism in the Combustor B can also exist in the two-phase rotating detonation system with high-temperature inlet air flow. Additionally, although a reaction cut-off line is artificially set in this study, the RDW can self-sustain far away from the cut-off line in the partially pre-evaporated cases. This indicates that the two types of combustion organization without and with IMS are both possible states in a real two-phase rotating detonation combustor. There may be specific formation and transition conditions between the two states, which can be discussed in the further studies.

**Acknowledgement**

The authors thanks to the support from the NSFC under the Grant No.51676111 and NSAF-U1730104.

**Appendix A: Validation of evaporation model**

Xu et al. [45] measured the evaporation behavior of a single $C_{10}H_{20}$ droplet at the ambient temperature of 633 K and ambient pressure of 1 atm. According to the experimental conditions, the evaporation process of three droplets are simulated. The initial diameters are 1.56, 1.23 and 0.92 mm, respectively. The variation of squared droplet diameter based on the experimental and numerical data are shown in Figure A1. The comparison shows that the evaporation model used in the study can accurately predict the droplet evaporation process.

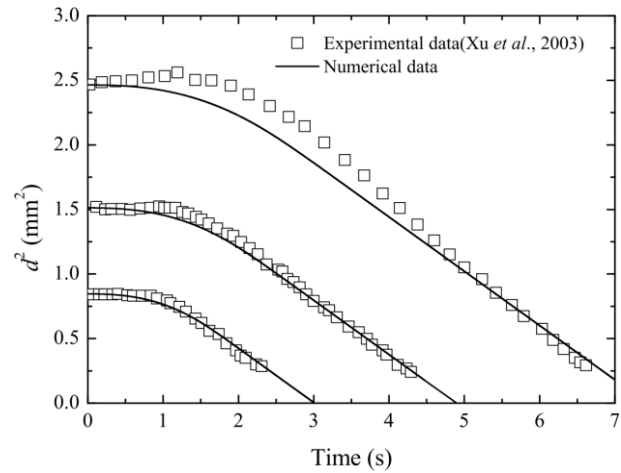

**Figure A1** Comparison of the droplet evaporation process in the experiments (Xu et al. [45]) and numerical simulations.

**Appendix B: Grid sensitivity analysis**

In order to check the grid independence, the simulation Case A5-A is recalculated with three grid sizes $\Delta x$ = 0.5, 0.25 and 0.125 mm, respectively. The time step size is chosen as $\Delta t = 5 \times 10^{-9}$ s. The pressure contours of the combustor with the three grid sizes are compared in Figure B1. The main structures of the flow field for three grid sizes are similar. The transverse waves on the detonation front cannot be captured with $\Delta x$ = 0.5 mm, while they are well simulated by $\Delta x$ = 0.25 and 0.125 mm. In addition, the average detonation wave velocities solved by $\Delta x$ = 0.5, 0.25 and 0.125 mm are very close, namely 1,723 m/s, 1,737 m/s and 1,740 m/s, respectively. Therefore, the grid size $\Delta x$ = 0.25 mm is chosen in this study.

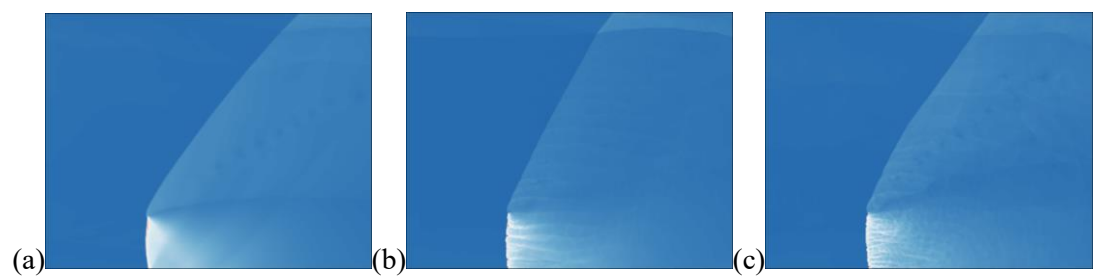

(a)  (b)  (c)

**Figure B1** Pressure contours of the combustor with different grid sizes $\Delta x$. (a) $\Delta x$ = 0.5mm; (b) $\Delta x$ = 0.25 mm; (c) $\Delta x$ = 0.125 mm.

**Reference**

[1] B. V Voitsekhovskii, Stationary spin detonation, Sov. J. Appl. Mech. Tech.